\documentclass[a4paper,11pt]{article}

\usepackage{graphicx}
\usepackage{amssymb}

\begin{document}

{\bf \LARGE Nucleation of single wall carbon nanotubes

of various chiralities}

\bigskip

Fran\c{c}ois Beuneu\hfill
\texttt{francois.beuneu@polytechnique.edu}

\bigskip

Laboratoire des Solides Irradi\'es, CNRS-CEA,
\'Ecole Polytechnique, F-91128 Palaiseau, France

\begin{abstract}
A simple model for the nucleation and growth of single wall carbon nanotubes from
a graphene sheet at the surface of a metallic catalyst saturated in carbon is developed. It enables
to predict the geometry and energy of tube embryos of all possible chiralities, as
well as the way that they can grow. It is shown that armchair-like chiralities are
energy preferred for geometrical reasons. This result is discussed and compared to experimental
literature.

\end{abstract}

\section{Introduction}
\label{sec:intro}

Since their discovery twenty years ago, carbon nanotubes (CNT) have attracted
a very strong interest among the scientific community. They are indeed fascinating objects
on an academic point of view, and they are foreseen for a large panel of future applications.
Among these potential applications, some rely strongly on the helicity of CNTs, mostly concerning electrical transport properties, but present
elaboration techniques do not enable to control this helicity: this is a challenge for the
future. It is highly desirable to better understand how CNTs with different helicities nucleate
and grow through their synthesis process, in the hope of controlling helicity.

In a preceding paper, I proposed a simple model of ``nucleation and growth of single wall
carbon nanotubes" \cite{beuneu05}. I showed that it is possible to build embryos of CNTs on
a graphene plane just by adding a number of bi-interstitials, and to make them grow by a similar
mechanism, involving a very restricted number of defects, namely carbon interstitials, Stone-Wales
defects and dislocations.

In the present paper, I continue this work, which was done systematically for a lot of geometries
and helicities, and I show that in the framework of the model some helicities are clearly
favored against others. I compare these results with experimental data, which are discussed.

\section{Model and calculations}
\label{sec:model}

The model for nucleation and growth of CNTs used here corresponds to the so-called
root-growth mechanism. A good example of the use of this mechanism is the vision
proposed by Maiti et al. \cite{maiti97}: a graphene layer wraps
metallic catalytic particles oversaturated in carbon. 
Embryos of CNTs, fed with interstitial C atoms dissolved in the catalyst, can form on this layer,
and subsequently grow from their foot.

More recent papers go in the same direction.
For instance, Gavillet et al. \cite{gavillet01} suggest a root-growth mechanism, after a study
both by high-resolution transmission electron microscopy and quantum-molecular-dynamics simulations.
They state that they ``will show that carbon atoms can be added at the root of a growing tube by a
diffusion-segregation process occurring at the surface of the catalytic particle."

Similarly, Ding et al. \cite{ding04}, in the framework of the VLS model (vapor-liquid-solid),
describe the tube growth ``by small graphitic islands that form on the supersaturated cluster surface."

The growth model suggested by Nasibulin et al. \cite{nasibulin05} is again quite similar.
A carbon layer is said to be present on the surface of the metallic catalyst particle.
In their fig.~5, they suggest that the C interstitials can come both
from the saturated catalytic grain and from carbon entering continuously the grain
through the opposite face of the particle. 

On this basis, my model lies on the fact that an embryo is nucleated in the graphene
plane, so that it can grow from the foot, perpendicularly to the
plane. Negative curvature on the graphene sheet
comes from the presence of heptagonal rings, and positive curvature from pentagonal rings.
The cap of the embryo must contain exactly 6 pentagons,
with exactly 6 heptagons around the foot of the embryo \cite{beuneu05}.

\begin{figure}
\includegraphics{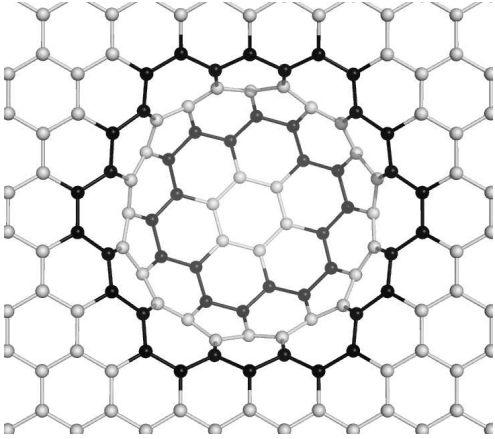}
\caption{\label{fig:build} Nucleation of an embryo of (12,0) CNT on a graphene plane:
24 C interstitials are added in the 12 cells delimited by the dark lines and atoms.}
\end{figure}

To build such an embryo, let us consider, on the honeycomb network of a graphene plane,
an hexagonal (regular or not) contour of elementary hexagonal cells, containing $N$
such hexagonal cells, $N$ being an integer greater or equal to 6.
We add $2N$ C interstitials, two per cell, coming from the carbon-saturated metallic catalyst.
An example of this process is given
in fig.~\ref{fig:build}: a 12-cell contour is delimited by the two dark lines,
24 C interstitials are added, and bounds are rebuild in such a way to get
a (12,0) zigzag tube. There is a lot of possibilities
to build bounds between the interstitials as well as between them and the other C atoms
of the plane. Among all these possibilities, there are several arrangements giving rise to
an embryo with the required 6 pentagons and 6 heptagons. Such arrangements can be labeled
by the usual indices $n$ and $m$, corresponding to a $(n,m)$ tube with $n+m=N$.
One can see quite easily that by such a process all the helicities, i.e. all the $n$
and $m$ values obeying $n+m=N$, are reachable; moreover, in general, there is a number
of isomers for each helicity. We can recall here the well-known fact that all helicities can
be described by taking both $m$ and $n$ as positive or zero integers, with the condition
$m \leq n$ (which implies $m \leq N/2$).

Once the interstitials are introduced and the bounds established, it is necessary
to let the bond length and angles relax towards equilibrium and to check
the stability and energy of the proposed defect geometries. I performed energy minimizations
by using the Tersoff model \cite{tersoff88}. The embryos are built on a parallelogram-shaped portion
of a graphene plane, with periodic conditions in the $x$ and $y$ directions. This gives rise
in fact to an embryo hexagonal network; I made calculations with varying the distance between
embryos, in terms of carbon hexagonal cell rows, from one to three rows. 

\section{Results}
\label{sec:results}

\begin{figure}
\hglue -15mm \includegraphics{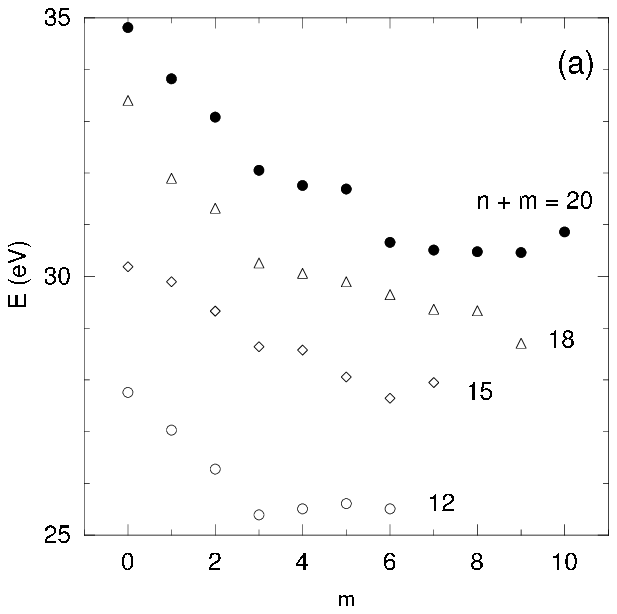}
\hglue 5mm \includegraphics{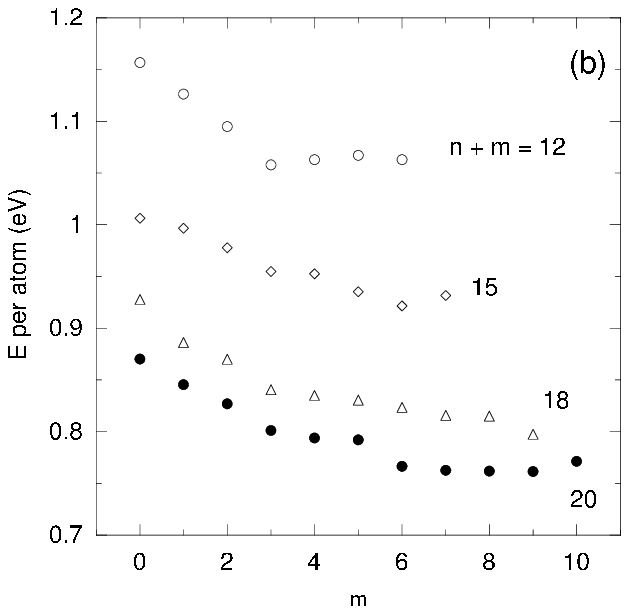}
\caption{\label{fig:edem} Energy of embryos of CNT versus $m$.
$N=n+m$ is 12 for white dots, 15 for diamonds, 18 for triangles
and 20 for black dots. (a): total energy; (b): energy per added atom.}
\end{figure}

I made calculations for embryos with $N=n+m$ equal to 12, 15, 18 and 20, and for all
possible helicities (from zigzag to armchair) in each case.
In the following, all energy values are given relatively to perfect graphene with
the same number of atoms.
On fig.~\ref{fig:edem}, I give the energy values for these geometries versus
the $m$ index, which is such that $m=0$ for zigzag tubes and $m=n$ for armchair ones.
Both total energy and energy per added atom values are given.

\begin{figure}
\includegraphics{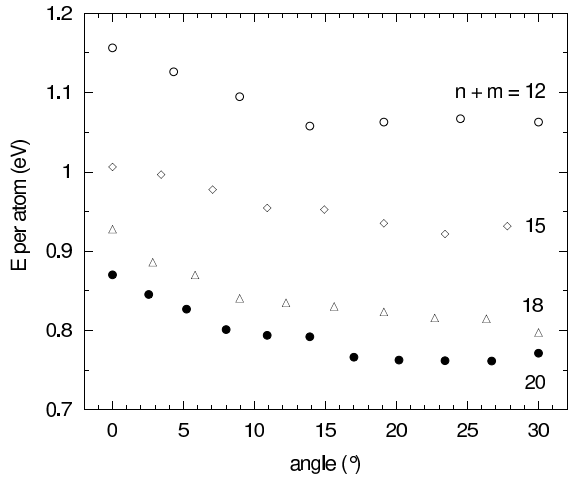}
\caption{\label{fig:edeangle} Energy of embryos of CNT versus chiral angle.}
\end{figure}

On fig.~\ref{fig:edeangle}, the same energy values are given versus
the chiral angle, whose value $\theta$ comes from the relation
$\theta=\arctan(\sqrt 3\times m/(m+2n))$.

\begin{figure}
\hglue -15mm \includegraphics{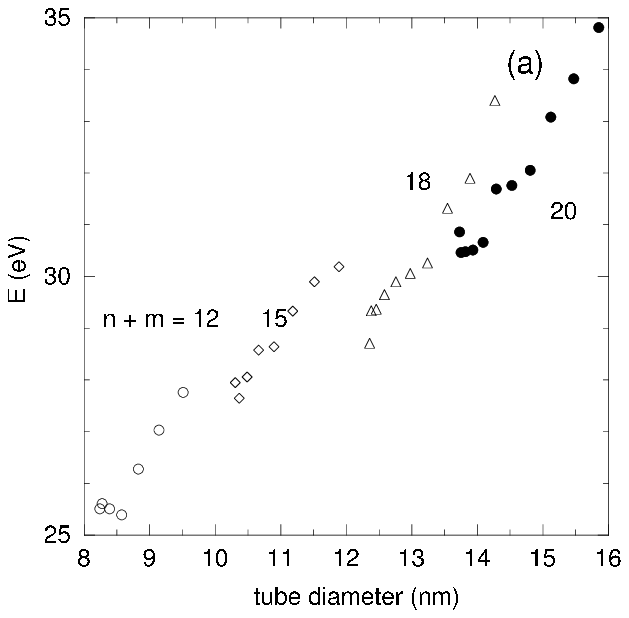}
\hglue 5mm \includegraphics{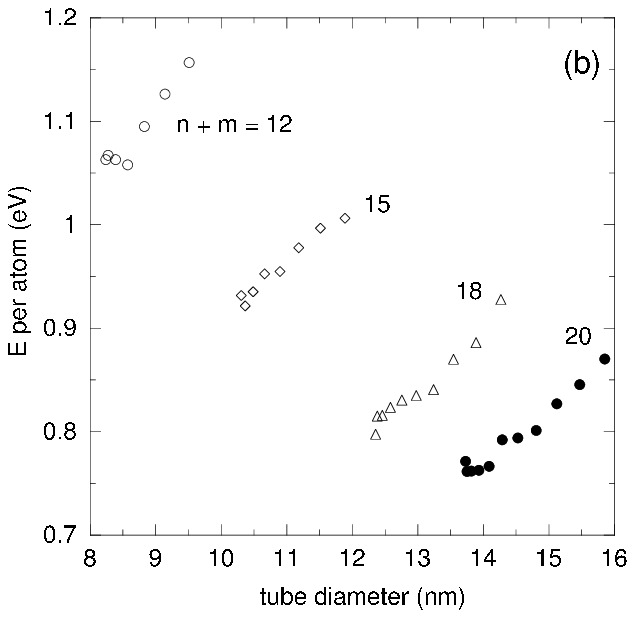}
\caption{\label{fig:edediam} Energy of embryos of CNT versus tube diameter.}
\end{figure}

On fig.~\ref{fig:edediam}, the same energy values are given versus
the theoretical tube diameter, whose value $D$ comes from the relation
$D=a \sqrt{n^2 + m^2 + nm}/\pi$ with the lattice parameter $a=0.249$~nm.

\begin{figure}
\includegraphics{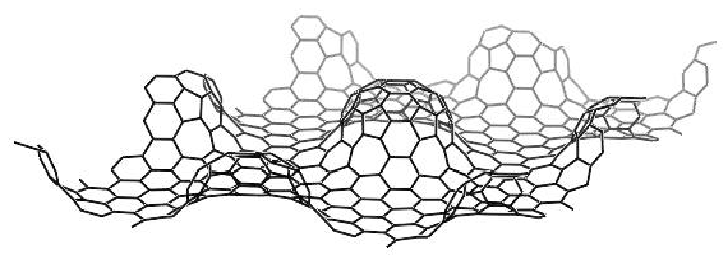}
\par
\includegraphics{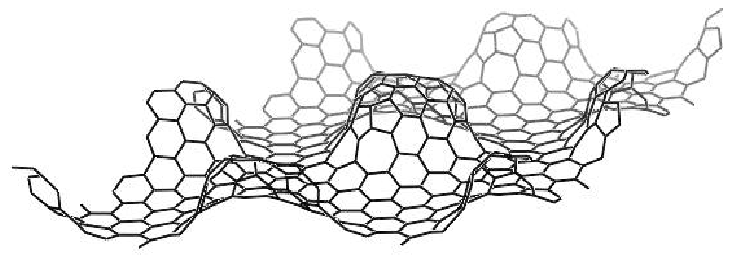}
\par
\includegraphics{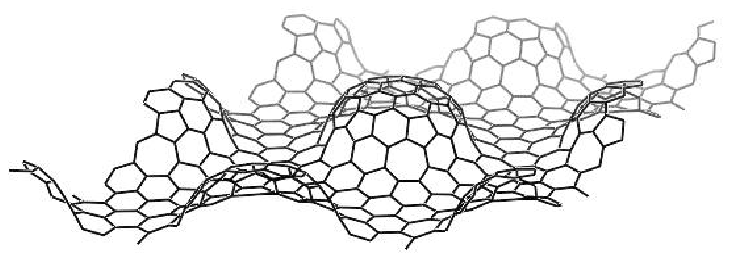}
\caption{\label{fig:chtidessin} Tube embryos with $N=12$ for three helicities,
from top to bottom: (6,6), (9,3) and (12,0).}
\end{figure}

For illustrating the geometry of the obtained embryos, I show on fig.~\ref{fig:chtidessin}
an example of tubes with $N=12$ for three helicities.

\begin{figure}
\hglue -15mm \includegraphics{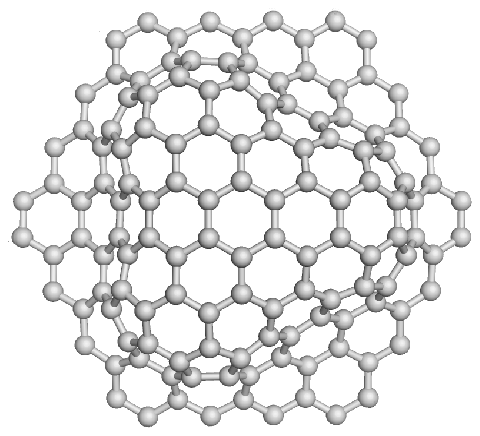}
\hglue -15mm \includegraphics{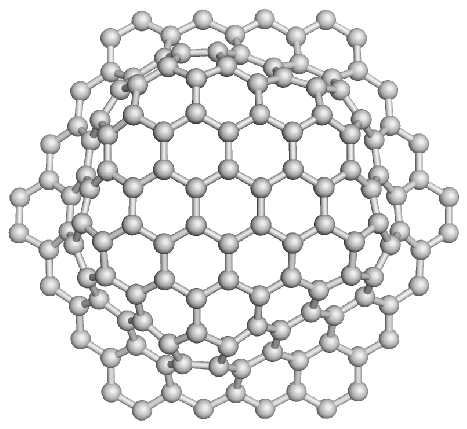}
\hglue 7mm \includegraphics{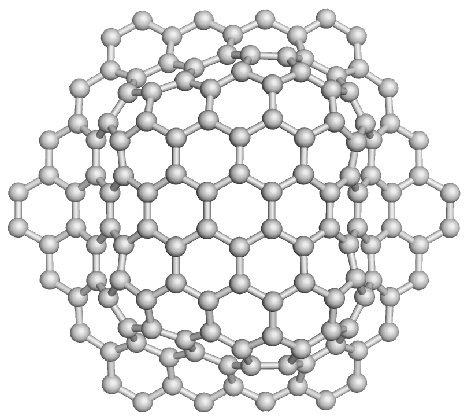}
\caption{\label{fig:isomeres} Nucleation of three isomers of an embryo of (9,9) armchair CNT
on a graphene plane: 36 C interstitials are added in the 18 cells delimited by the external lines.
The 6 heptagons are situated at different places on the sides of the hexagonal contour.}
\end{figure}

I mentioned above, in section \ref{sec:model}, that there is in general
a substantial number of isomers for each helicity. This is especially true for values of $n+m=N$
larger than 12, when there are several possible configurations for the 6 heptagons in the
hexagonal contour of the embryo. In practice, there is a small distribution
of energies for the different isomers of a given $(n,m)$ embryo, of the order of 1~eV.
One notes that for zigzag embryos, there is only one isomer, when the 6 heptagons are situated
at the vertices of the hexagonal contour; this is true for all sizes of zigzag embryos,
and does not depend on the fact that the contour hexagon is regular or not.
For all other helicity cases, different isomers exist in general,
corresponding to different possible repartitions of heptagons on the vertices or sides
of the contour. An example is shown in fig.~\ref{fig:isomeres} for an embryo
of a (9,9) armchair nanotube, with three possible configurations of the heptagons on the sides.

\begin{figure}
\includegraphics{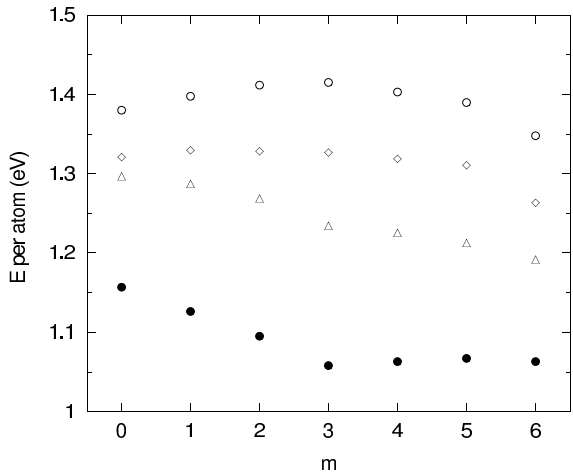}
\caption{\label{fig:croissance} Growth of $N=12$ CNTs.
black dots: initial embryo; triangles: one row added;
diamonds: two rows added; white dots: three rows added.}
\end{figure}

Starting from the calculated embryos, it is easy to grow the CNTs, row by row. For each
row, it is necessary to add between the foot and the cap $2N$ carbon bi-interstitials.
I did such calculations, up to three rows, and got the energy values corresponding to the obtained
geometry after the Tersoff minimization. An example is given fig.~\ref{fig:croissance}.

\section{Discussion}
\label{sec:discussion}

It seems very obvious from fig.~\ref{fig:edem} and \ref{fig:edeangle} that the energy
of embryos is always maximum for $m=0$ (zigzag tubes) and is minimum for values of $m$
close to $n$, that is tubes close to the armchair geometry. The energy difference between
the extreme geometries is about 2 to 4~eV, corresponding roughly to 0.1~eV per C interstitial.
The fact that there are several isomers in the armchair geometry whereas there is only one
for zigzag tubes goes in the same direction: armchair or quasi-armchair tubes are predicted to
be favorised in such a nucleation and growth process.

There are several papers in literature dealing with CNT production and chirality measurement.
I review now some of these papers, showing that the conclusions
are quite far from unanimity.
Cowley et al. \cite{cowley97} observed in CNT ropes produced by laser irradiation
that the predominant helicity was armchair - (10,10) - or close to it.
Qin et al. \cite{qin97} found, also in laser-produced tubes, that in most cases the helical angles
were evenly distributed, except a few cases where armchair structure was preferred.
Bernaerts et al. \cite{bernaerts98} worked also on laser tubes and concluded that the ropes
comprised largely tubes of armchair chirality, but not exclusively (10,10).
Henrard et al. \cite{henrard00} found that the situation was somewhat more complex; pointing the disagreement between the preceding papers, they claimed that within a given bundle
no chirality was favoured.
Colomer et al. \cite{colomer01,colomer02} worked on small bundles produced by catalytic chemical
vapor deposition (CVD) and claimed that except for a weak tendency for the ``armchair structure",
no preferred helicity was found.
Bachilo et al. \cite{bachilo02} observed a lot of semiconducting tubes grown
in high-pressure carbon monoxide (HiPco) and obtained a random distribution of helicities;
in a subsequent paper \cite{bachilo03}, they analyzed also tubes obtained by CVD and found
then a significantly stronger preference for near-armchair structures.
Li et al. \cite{li04,li05} published a strong preferential growth of semiconducting tubes
in the case of a plasma enhanced CVD technique, very different than in the HiPco or laser
processes.
Gavillet et al. \cite{gavillet04} concluded that ``the helicities of the different tubes
in a bundle obtained from the arc and the laser methods are distributed at random" and
suggested that in the case of CVD, some bundles contain tubes with identical helicities.
Vitali et al. \cite{vitali04}, in the case of HiPco tubes, did not find any preferential
helicity in the tubes that they analyzed.
Miyauchi et al. \cite{miyauchi04} studied tubes synthesized from the alcohol catalytic CVD method
and claimed that chiralities were distributed predominantly close to the armchair structure.
Meyer et al. \cite{meyer05}, working on CVD tubes, also found quite dispersed helicities.
It is worth mentioning the extended work of Liu et al. \cite{liu05} who studied
124 single-walled CNTs synthesized by the arc method, and noticed a slight preference
for the helicity ${15^\circ}$-${30^\circ}$, quite close to the armchair structure.

Reich et al. \cite{reich05,reich06} gave a comprehensive theoretical study of the nucleation
and growth of CNT caps on a catalyst surface. They pointed the fact that in a root-growth
mechanism, the initial geometry of the cap controls the tube chirality. In such a framework,
the epitaxial relation between the cap and the catalyst surface is of major importance for
controlling chirality.

An interesting work is the one by Lolli et al. \cite{lolli06}. They proposed a way
to vary the distribution of $(n,m)$ tubes with high specificity, by using CoMo catalysts
and two C-rich gases, CO and CH$_4$. With CO feed, they got rather small tube diameters,
with a strong preference to the armchair or quasi-armchair geometry, like Miyauchi et al.
With CH$_4$, by contrast, tubes appeared near both the zigzag and armchair lines.

In a recently published book, Page et al. \cite{page11} gave an extensive review
on the mechanisms of single-walled CNT nucleation, growth and chirality-control,
mainly focused on quantum-mechanical molecular dynamics simulations. In the experimental
part of their review, they described not only the classical techniques like arc discharge or
CVD but also more non-traditional methods using new catalysts such as SiO$_2$, SiC
and Al$_2$O$_3$.

Finally, Sankaran \cite{sankaran11} presented the recent advances and potential of plasma
technology for CNT chirality control. He pointed that the catalyst structure can influence
the nucleation of specific chiralities. Varying the plasma process temperature could also
be a way to controlling the chirality.

It is hard to deduce a precise tendency from this literature: for the different CNT
growing techniques, is there a preferential nucleation of some helicities? The cited
papers are somewhat contradictory on that question. However, it seems to be an
overall preference for armchair-type CNTs; see for instance
\cite{cowley97,bernaerts98,bachilo03,miyauchi04} and particularly \cite{lolli06}.
This tendency to get CNTs more on the armchair structure side is coherent with
the findings of the calculations exposed in the present paper.

I wish to add a few words about energy calculations. Li et al. \cite[fig~4a]{li05}
show density functional theory calculations of the formation energy of various CNTs
versus tube diameter $d$: for tubes without foot or cap, they get a $1/d^2$ variation;
for $d=1$ nm, $E$ is roughly 0.1 eV/atom. They also notice
a small preference for semiconducting tubes.
Liu et al. \cite{liu05}, using ab initio calculations, found less than 20~meV/at
of energy difference between the different CNT structures.
Reich et al. \cite[table I]{reich05} gave a lot of ab initio data on (10,0)-tube
energies: 0.137 eV/at for the tube without cap, and between 0.29 and 0.37 eV/at
for the tube cap. My own calculations give 0.08 eV/at for a (12,0)-tube growth
(fig.~\ref{fig:croissance}) and 0.43 eV/at for the same tube cap nucleation.

\section{Conclusion}
\label{sec:conclusion}

The present paper develops the calculations for nucleation and growth of CNTs
from a graphene sheet at the surface of a catalyst. The geometrical constraints
of this process induce some energy differences between the different tube geometries,
which are clearly in favor of armchair or quasi-armchair tube structures.

There is no totally unambiguous experimental evidence of such a preference in the classical
tube growth methods. However, several studies notice such a tendency to nucleate preferentially
tubes of armchair-type chirality.



\end{document}